# Nonequilibrium fluctuations of global warming

Jun Yin[1], Amilcare Porporato[2*] and Lamberto Rondoni[3,4]

[1]Department of Hydrometeorology, Nanjing University of Information Science and Technology, Nanjing, 210044, China.
[2]Department of Civil and Environmental Engineering, Princeton University, Princeton, 08540, USA.
[3]Dipartimento di Scienze Matematiche, Politecnico di Torino, 10125 Turin, Italy
[4]INFN, Sezione di Torino, Via P. Giuria 1, 10125 Torino, Italy

Corresponding author: aporpora@princeton.edu

**Abstract**

While the warming trends of the Earth's mean temperature are evident at climatological scales, the local temperature at shorter timescales are highly fluctuating. In this letter we show that the probabilities of such fluctuations are characterized by a special symmetry typical of systems out of equilibrium. Their nearly universal properties are linked to the fluctuation theorem and reveal that the progressive warming is accompanied by growing asymmetry of temperature distributions. These statistics allow us to project the global temperature variability in the near future, in line with predictions from climate models, providing valuable information about future extremes.

The anthropogenic increase in greenhouse gases can be seen as a thermodynamic protocol driving the system toward new nonequilibrium conditions via modifications of the system's properties. These changes are altering ocean and atmospheric currents, the hydrologic and biogeochemical cycles, threatening ecosystems and human life (Doney et al. 2012; Grimm et al. 2013). However, the fact that the climate system is a nonlinear open system, inherently chaotic in both space and time (Bohr and van de Water 1994; Cross and Hohenberg 1994) makes the understanding of the multiscale nature of the Earth's climatic fluctuations particularly challenging.

The overall warming trends are embedded within the warming and cooling patterns controlled by the large-scale circulation (e.g., at synoptic and interannual scales) (Lu et al. 2007; Seager et al. 2010). Climate change tends to be more evident in processes that integrate temperature variability at long timescales, such as glacial melting and sea level rise (Tebaldi et al. 2021), while it may be somewhat masked by fluctuations when focusing on quantities, such as temperature and precipitation extremes, which vary more erratically (Hansen et al. 2012; Trenberth et al. 2015). Even the mean annual temperature, which is typically used as a measure of climate change progression, may be deceiving due to its strong space-time variability: for example, in 2020 – one of the hottest years on record – nearly half (49.1%) of the Earth was actually colder than the previous year (see Fig. 1a).

The recent advances in statistical mechanics on the structure of fluctuations of nonequilibrium phenomena (Evans et al. 1993; Gallavotti and Cohen 1995; Crooks 1999; Evans and Searles 2002; Jarzynski 2004; Marconi et al. 2008; Evans and Morriss 2008; Seifert 2012; Gallavotti 2014) may be helpful to characterize the structure of the Earth's global warming fluctuations. Initially concerning the entropy production of chaotic many particle systems, such theories have derived novel symmetries that point to a certain degree of universality in the nonequilibrium fluctuations of observables and in the response to perturbations of large as well as "small" systems. In particular, small systems are characterized by fluctuations of size comparable to the observed average signals, hence they cannot be neglected. A deeper understanding of the macroscopic properties of such systems has emerged from leveraging the statistics of instantaneous microscopic events, which often look totally random and *per se* do not appear to have a direction of time but, when averaged in time, they reveal a preferred direction for the overall evolution. This can be used to quantify the degree of irreversibility of the system, and to identify an asymptotic universal behavior verifying scaling relationships that help extrapolate it to long times.

Fluctuation relations can be verified when a macroscopic object is observed on a microscopic scale (e.g., (Bonaldi et al. 2009)). However, the theory of nonequilibrium fluctuations is of particular interest in the case of systems with constituting elements interacting on scales comparable with those of the overall system, because fluctuations are then *observable* on the same scale of the global phenomenon. Dissipative complex phenomena such as turbulence (Ciliberto and Laroche 1998; Gallavotti et al. 2004; Porporato et al. 2020) and the behavior of bacteria (Bechinger et al. 2016) have been shown to follow similar rules. This is consistent with the small system picture, because the relevant elementary constituents (eddies or single bacteria) are moderately numerous and interact on the scale of the whole system. The ensuing response theory of both near- and far-from-equilibrium fluctuations (Ghil and Lucarini 2020) is a promising framework to analyze the macroscopic patterns of global warming and shed light into the Earth's climate dynamics. The hydroclimatic fluctuations in the Earth's climate are in fact also comparable with the Earth's size, effectively rendering its behavior like the one of small systems from the point of view of statistical mechanics. According to this view, the Earth is driven out of equilibrium by input of low entropy radiation and an outgoing high-entropy longwave radiation (Kleidon 2016).

We consider the spatial patterns of global temperature anomalies at different temporal scales. As shown in the global maps of Fig. 1, when focusing on short time scales, nearly half of the Earth has positive anomaly differences (black solid lines in Fig. 1a) and the spatial distribution of warmer or cooler regions changes almost randomly (e.g., Fig. 1c). Even for the North Pole, one of the fastest warming regions (Bekryaev et al. 2010; Stuecker et al. 2018), still only half the region is getting warmer each year (see Fig. S1). However, as the averaging window and time difference are increased, a significant portion of the Earth appears to be getting warmer (yellow dash-dot lines in Fig. 1a and spatial patterns in Fig. 1c). Similar results are obtained from other climate data (combined sea surface temperature over the ocean and near-surface air temperature over the land; see Fig. S2). The warming signal becomes stronger with longer timescales (see Fig. 1b) as the fluctuations get averaged out.

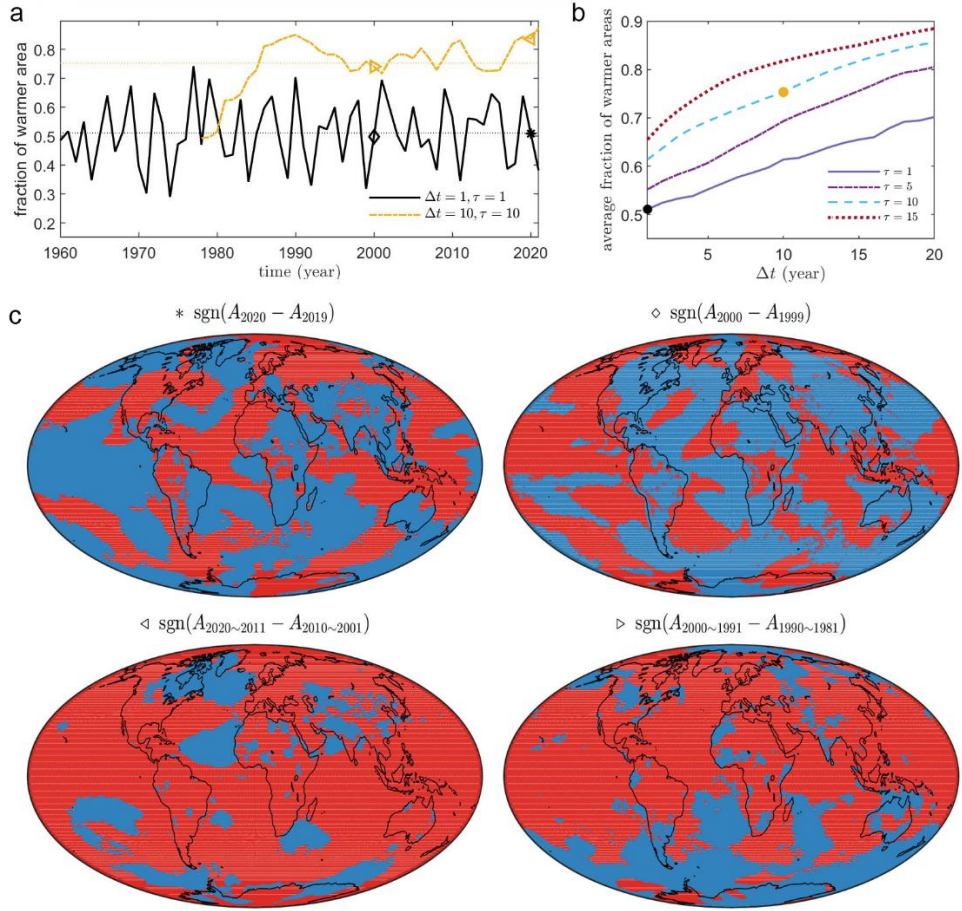

Fig. 1 Multiscale analysis of global warming fluctuations. (**a**) Fraction of the Earth's surface having annual temperatures higher than the previous year (black solid lines) or decade-average temperatures higher than the previous decade (yellow dash-dot lines). (**b**)

Average fraction of warmer areas with the varying time difference ($\Delta t$) and averaging windows ($\tau$). The black and yellow dots in (b) correspond to the black and yellow horizontal dash lines in (a) respectively. (**c**) Geographical distributions of warmer (red) and colder (blue) regions with different averaging windows and time differences. The areal fraction of the total warmer regions in (c) are marked in (a) with the same star, diamond, and left/right triangles. Results are based on ERA-5; see similar patterns for other data sources in Fig. S2.

To quantify the asymmetry of these temperature fluctuations, we consider the temperature anomalies, *a*, defined as the deviation from the average temperature over 1959-2014, covering a period with available data from both reanalysis and climate model outputs. The choice of base period does not change the asymmetric patterns discovered in this study (see Fig. S9 for results based on different base periods). The local anomaly is averaged over a time window $\tau$

$$A(t;\tau) = \frac{1}{\tau} \int_{t-\tau}^{t} a(t)dt , \quad (1)$$

and also compared between two periods separated by a time difference $\Delta t$

$$\Delta A(t;\tau, \Delta t) = A(t;\tau) - A(t-\Delta t;\tau) . \quad (2)$$

The results show that the probability density functions (PDFs) of both *A* and $\Delta A$ exhibit a remarkable structure, which is robust when changing averaging and differencing time windows, consistently with the theory of nonequilibrium fluctuations. As shown in Fig. 2 a and c, the obtained PDF consistently presents double exponential tails (straight lines in the logarithmic plots), which become more evident for longer averaging windows. The shape of the distributions is well captured by asymmetric double exponential distributions (Kozubowski and Podgórski 2000; Yu and Zhang 2005), with probability density function (PDF)

$$f(x;m,\beta_1,\beta_2) = \begin{cases} \beta_0 \exp[-(x-m)/\beta_1] & x > m \\ \beta_0 \exp[(x-m)/\beta_2] & x \leq m \end{cases} \quad (3)$$

where the random variable *x* may refer to either *A* or $\Delta A$, *m* is the mode, $\beta_1$ and $\beta_2$ are the exponents for the right and left tails, and $\beta_0 = 1/(\beta_1 + \beta_2)$.

Interestingly, the double exponential shape gradually turns anticlockwise, giving rise to a positive skewness in the more recent years (darker lines in Fig. 2 a and c). This reveals a warming trend toward more extreme events, as well as an acceleration of global warming pace. The change in the PDF properties is summarized by the plots of temperature averages against the standard deviation and the asymmetry index, $\Delta\beta = \beta_1 - \beta_2$, a measure of asymmetry of the two tails derived from the fluctuation relations. As shown in Fig. 2 b and d, the variance is smaller for longer averaging windows and has only a weak dependence on the mean, first decreasing and then increasing for $A$ and gradually increasing for $\Delta A$, whereas the distributions become more asymmetric as the mean increases; again such patterns become more evident for larger averaging windows.

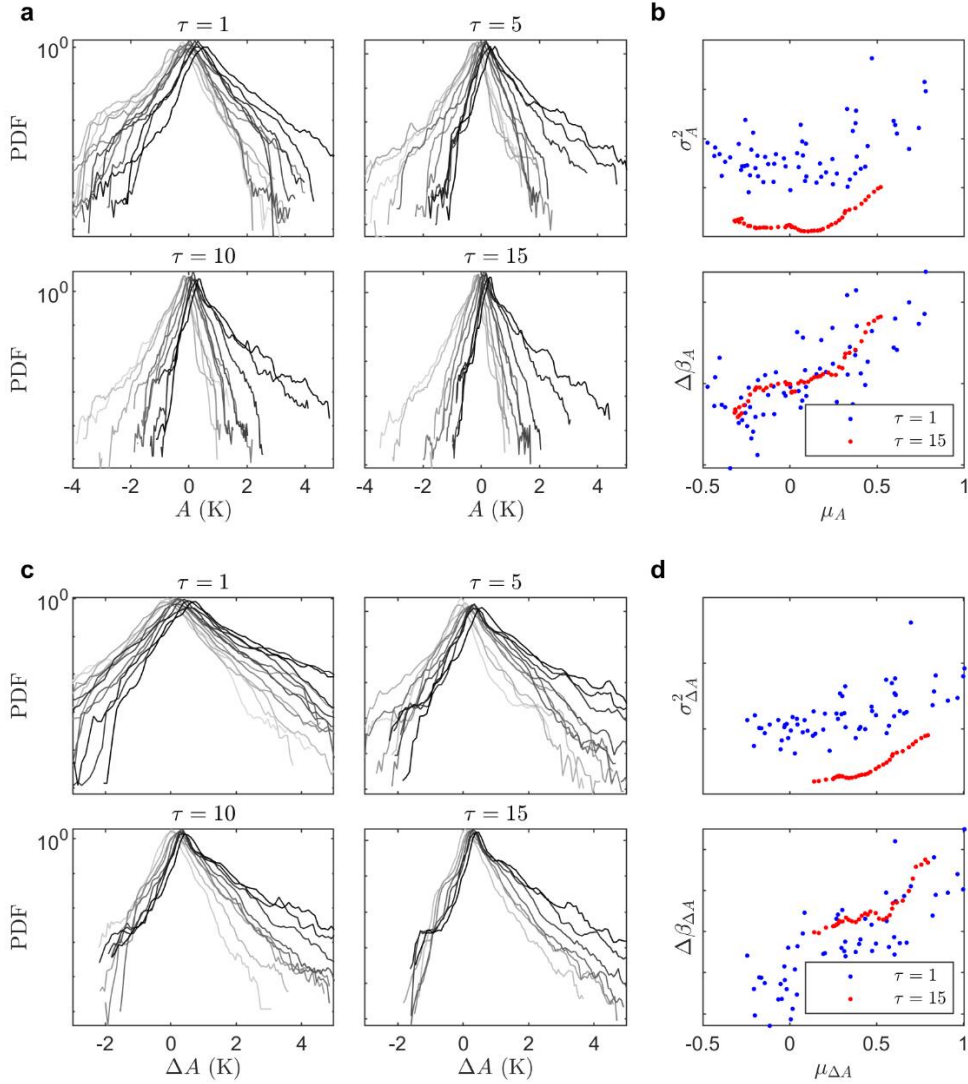

Fig. 2 Global distributions and statistics of temperature fluctuations. Probability density functions (PDFs) of anomalies (**a**, $A$) and anomaly differences (**c**, $\Delta A$) with different averaging windows ($\tau$) are displayed in chronological order with lighter colors for earlier years (or shorter time differences) and darker colors for the later ones (or longer

time differences). The baseline for $\Delta A$ is set as the earliest year of the data series. The relationships among mean $\mu$, variances $\sigma^2$, and asymmetry $\Delta \beta$ of the anomalies (**b**, $A$) and anomaly differences (**d**, $\Delta A$) are presented in blue dots for the short averaging window ($\tau = 1$ year) and in red dots for the long averaging window ($\tau = 15$ years). Results are based on ERA-5; see similar patterns for other data sources in Fig. S3.

As required by fluctuation relations, which concern the asymptotic time averages of fluctuations, we rescaled both sides of the distributions by the corresponding exponents, i.e.,

$$\bar{x} = \begin{cases} (x-m)/\beta_1, & x > m \\ (x-m)/\beta_2, & x \leq m, \end{cases} \quad (4)$$

where the overbar refers to the scaled variable. After this scaling, the theoretical distribution becomes

$$f(\bar{x}) = \beta_0 \exp(-|\bar{x}|), \quad (5)$$

which has a unit exponent for both tails. The resulting PDFs from data are shown in Fig. 3a and 3c, confirming the almost complete collapse onto a single curve. This allows us to accurately project the global temperature patterns as functions of only three parameters (the two exponents and the mode).

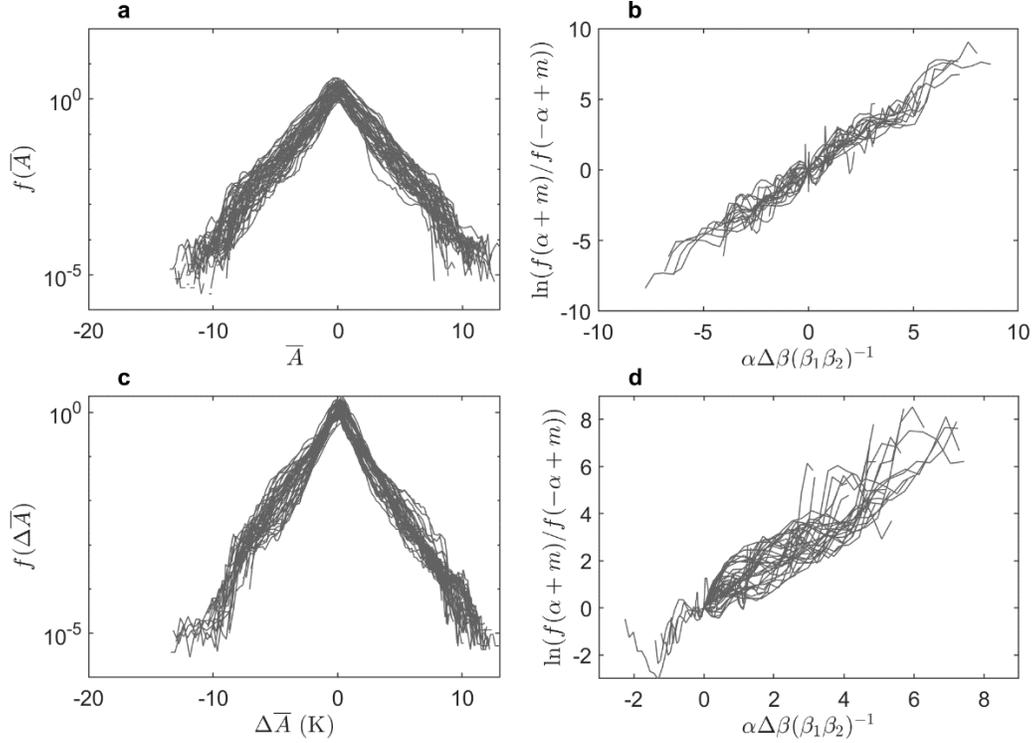

Fig. 3. Rescaling and symmetry of temperature fluctuations. The lines in (**a**) and (**c**) are rescaled from the PDFs of $A$ and $\Delta A$ in Fig. 2 a and c, respectively. The tails of these PDFs are linearly stretched to have a unit slope (see Materials and Methods). The symmetry of the temperature fluctuations is evaluated by the ratio of positive and negative temperature fluctuations away from the mode against the differences of the exponents of the tails for A (**b**) and for $\Delta A$ (**d**). Results are based on ERA-5; see similar patterns for other data sources in Fig. S6.

The asymmetry of the distribution with the growth of positive fluctuations at the expense of the negative ones is strongly reminiscent of the symmetry implied by the fluctuation theorem (Evans et al. 1993; Marconi et al. 2008), whereby the ratio of the probabilities taken on opposite signs is related to the dissipation and hence the irreversibility of the nonequilibrium process. As appropriate for large deviation results, we considered the logarithm of the ratio of positive and negative temperature fluctuations away from the mode in a distance of $\alpha$,

$$\ln\left[\frac{f(\alpha+m)}{f(-\alpha+m)}\right] = \alpha\Delta\beta(\beta_1\beta_2)^{-1}, \qquad (6)$$

which turns out to be related to the exponents of the two tails, as one would have predicted from the symmetry of the fluctuation theorem. The results, shown in Fig. 3 b and d, synthesize the link between the cooling and warming probabilities and the overall warming trend. Its warming-to-cooling ratio, governed by Eq. (7), increases for stronger asymmetric distribution (e.g., longer time differences and more recent years).

The robust scaling of the PDFs and its shift in asymmetry may also be used to extrapolate the data for short lead times to infer global warming trends (see Fig. 4a-c). This was done by fitting the statistics of *A* with smoothing window $\tau = 15$ to the year of 2040. A linear function was first chosen to fit the data but a quadratic function was used to capture the nonlinear behaviors of the variance (see triangles in Fig. 4a-c). This by no means covers all possible extrapolations, but provides a first glimpse into future conditions. With the projected mean, variance, and asymmetry index, we can obtain all parameters for the asymmetric double exponential distribution (i.e., $m$, $\beta_1$, $\beta_2$) and provide future distributions of temperature anomaly (see dash lines in Fig. 4d).

Consistent with climate model projections (Masson-Delmotte et al. 2021), the extrapolation shows that a 1.5-degree mean temperature increase is reached at around 2040, along with an increase in variance and asymmetry. With these statistics, we obtain a distribution of future temperature anomaly that is extremely skewed, with a shortened left tail and an elongated right tail (see dash lines in Fig. 4d). When compared with climate model outputs (magenta shading and lines) and the observations in recent years, similar patterns are observed, but with a slightly faster increase of mean temperature and a slower increase of skewness (also see alternative projections in Fig. S7). The simulated temperature distribution is consistent with our extrapolation results with almost identical right tails, revealing the goodness of the fluctuation approach and corroborating our finding of extreme warming trends in the near future.

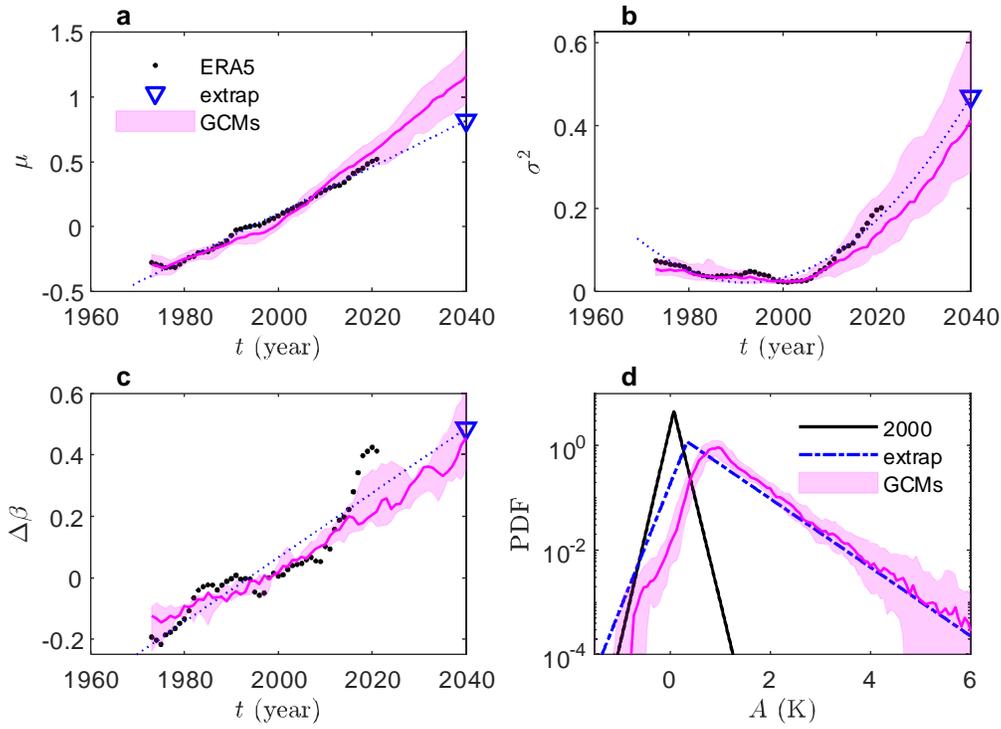

Fig. 4 Future projection of temperature fluctuation. Time series of (**a**) mean, (**b**) variance, and (**c**) asymmetry index of temperature anomalies with an averaging window of 15 years from ERA-5 data (thick black dots), compared with those from climate model outputs (magenta shading and lines). Triangles are extrapolation using ERA-5 data; results from climate model outputs before and after 2014 were from 'historical' and 'ssp245' experiments, respectively. A quadratic function was used to capture the nonlinear shape of the variance, whereas linear functions were used for other extrapolations. (**d**) Distributions of temperature anomalies from climate model outputs (magenta shading and lines) are compared with the asymmetric double exponential distributions, where the mean, variance, and symmetric index were from values in 2000 (black lines) and the extrapolated values in 2040 (dash lines). The magenta shading marks to the 25th and 75th percentiles of the model outputs and magenta lines show the median values (see Table S1 for model list); results from each climate model are reported in Fig. S8.

The exponential tails are not uncommon for many atmospheric variables, whose time series in certain regions may show non-Gaussian distributions (Perron and Sura 2013; Proistosescu et al. 2016; Catalano et al. 2020). Such fat tails may be associated with advection-diffusion process (Pierrehumbert 2000; Neelin et al. 2010) and have important implications for future occurrences of the extreme events (Ruff and Neelin 2012). The spatial distributions and its asymmetric shift in time explored here not only corroborate the higher probabilities of extreme events in a changing climate but also indicate their continued existence with almost random appearance over the world. The implications of such spatiotemporal distributions, and the applicability of fluctuation relations, for extreme events deserve closer attention, in view of their impact on the large-scale circulation, the hydrological and biogeochemical cycles, as well as ecosystems and society (Easterling et al. 2000; van de Pol et al. 2017), and on the understanding of such phenomena.

In summary, our analysis based on the fluctuation relation discovered thirty years ago, uncovered a remarkable symmetry of global temperature fluctuations, which stems from the behavior of the Earth's climate as a 'small' nonequilibrium system. This in turn allowed us to collapse space-time temperature fluctuations into a single scaling law, resulting in a universal shape of the distribution of the temperature anomalies. When extrapolating to the near future, the temperature anomalies present a skewed distribution with elongated tails, highlighting the frequent occurrence of extreme events. The scaling law uncovered here may serve as the benchmark for climate model simulations as well as to infer the Earth's response to the changing climates. Future work will be devoted to extending this type of analysis to other important variables including precipitation and energy-cycle statistics with the goal of harnessing the formalism of response theory to improve our understanding, simulation, and prediction of the Earth system.

**Acknowledgements**

J.Y. acknowledges support from the Natural Science Foundation of Jiangsu Province (BK20221343). A.P. acknowledges support from the Carbon Mitigation Initiative at Princeton. L.R. acknowledges financial support by the Ministero dell'Università e della Ricerca (Italy), grant Dipartimenti di Eccellenza 2018–2022 (E11G18000350001). His research was performed under the auspices of Italian National Group of Mathematical Physics (GNFM) of INdAM.